\documentclass{iaus}
\usepackage{graphicx}

\title[Cosmic-ray driven dynamo in the medium of irregular galaxies] 
{Cosmic-ray driven dynamo in the medium of irregular galaxy}

\author[H. Siejkowski et al.]   
{Hubert Siejkowski$^1$, Marian Soida$^1$, Katarzyna Otmianowska-Mazur$^1$, Micha\l{} Hanasz$^2$, Dominik Bomans$^3$}

\affiliation{
$^1$Astronomical Observatory, Jagiellonian University, ul. Orla 171, 30-244 Krak\'ow, Poland \break email: h.siejkowski@oa.uj.edu.pl\\[\affilskip]
$^2$Toru\'n Centre for Astronomy, Nicolaus Copernicus University, 87-148 Toru\'n/Piwnice, Poland\\[\affilskip]
$^3$Astronomical Institute of Ruhr-University Bochum, Univerist\"{a}tsstr. 150/NA7, D-44780 Bochum, Germany}

\pubyear{2009}
\volume{259}  
\pagerange{100--100}
\date{"YOUR MAILING DATE"  and in revised form ??}
 \jname{Cosmic Magnetic Fields: From Planets,
to Stars and Galaxies} \editors{K.G. Strassmeier, A.G. Kosovichev \&
J.E. Beckman, eds.}
\begin{document}

\maketitle

\begin{abstract}
We investigate the cosmic ray driven dynamo in the interstellar medium of irregular galaxy.  The observations
(Chy\.zy et al. 2000, 2003) show that the magnetic field in irregular galaxies is present and its value
reaches the same level as in spiral galaxies. However the conditions in the medium of irregular galaxy are
very unfavorable for amplification the magnetic field due to slow rotation and low shearing rate.

In this work we present numerical model of the interstellar medium in irregular galaxies. The model includes
magnetohydrodynamical dynamo driven by cosmic rays in the interstellar medium provided by random supernova
explosions. We describe models characterized by different shear and rotation. We find that even slow galactic
rotation with low shearing rate gives amplification of the magnetic field. Simulations have shown that high
amount of the magnetic energy flow out off the simulation region becoming an efficient source of intergalactic
magnetic fields.

\end{abstract}

\firstsection 


\section{Model and input parameters}
The CR-driven dynamo model consists of the following elements (based on Hanasz et al. 2004, 2006): (1) the
cosmic ray component is an relativistic gas described by diffusion-advection transport equation; (2)
anisotropic diffusion of CR along magnetic field lines; (3) localized sources of CR: random explosions of
supernova remnants in the disk volume, which supply the CR energy density; (4) uniform resistivity of ISM; (5)
shearing boundary conditions and tidal forces are implemented to reproduce the differentially rotating disk in
the local approximation (Hawley et al. 1995); (6) realistic vertical disk gravity following the model of Milky
Way (Ferri\'ere 1998) with scaled contribution of disk and halo to irregular galaxies. We used resistive MHD equations in 3D
Cartesian domain of $25\times50\times400$ grid points. The initial conditions of system assumes the
magnetohydrostatic equilibrium with horizontal, purely azimuthal magnetic filed corresponding to $p_{\it
mag}/p_{\it gas} = 10^{-4}$.


\section{Results}

\begin{figure}[t]
\centering
\includegraphics[width=6.5cm]{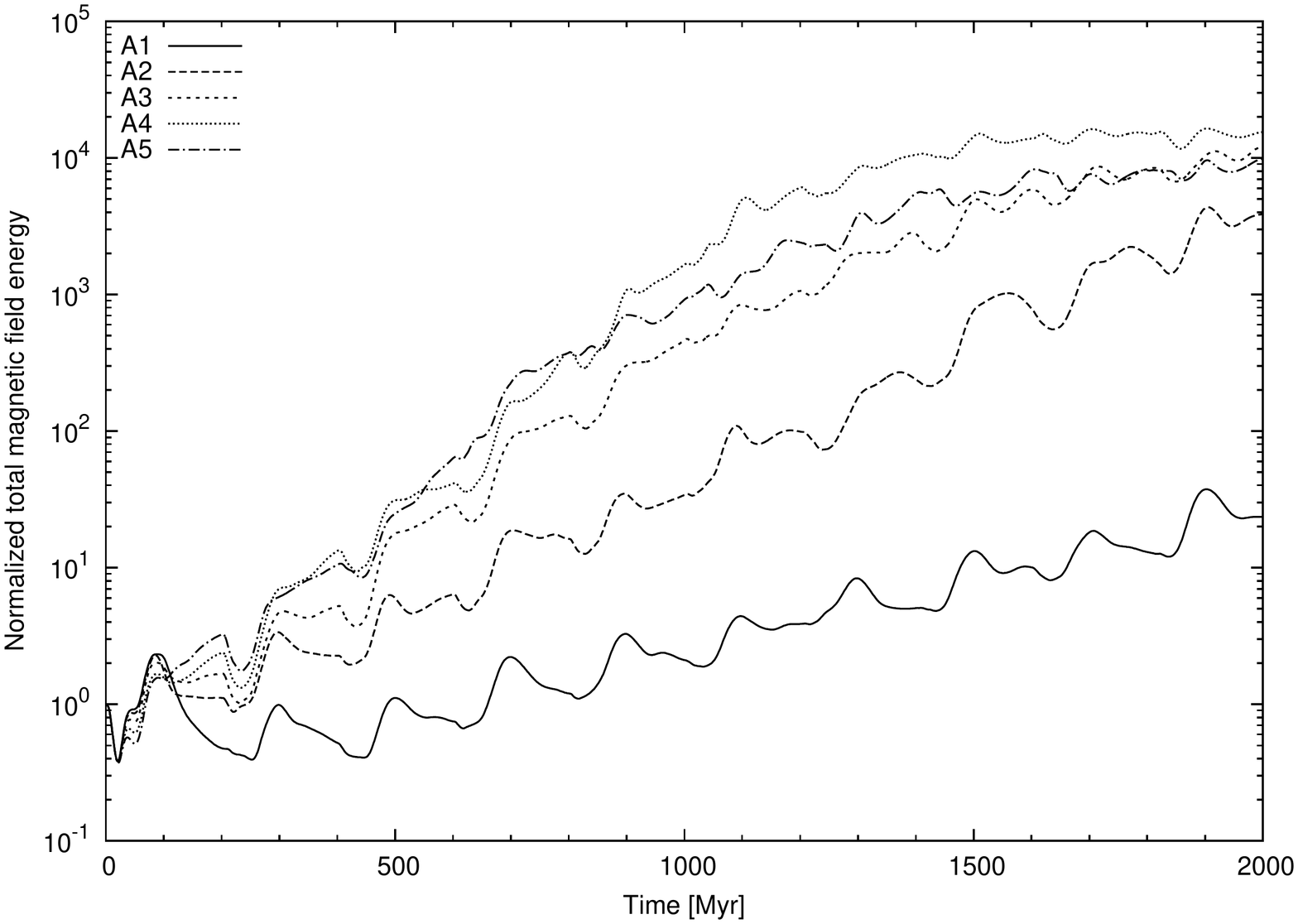}
\includegraphics[width=6.5cm]{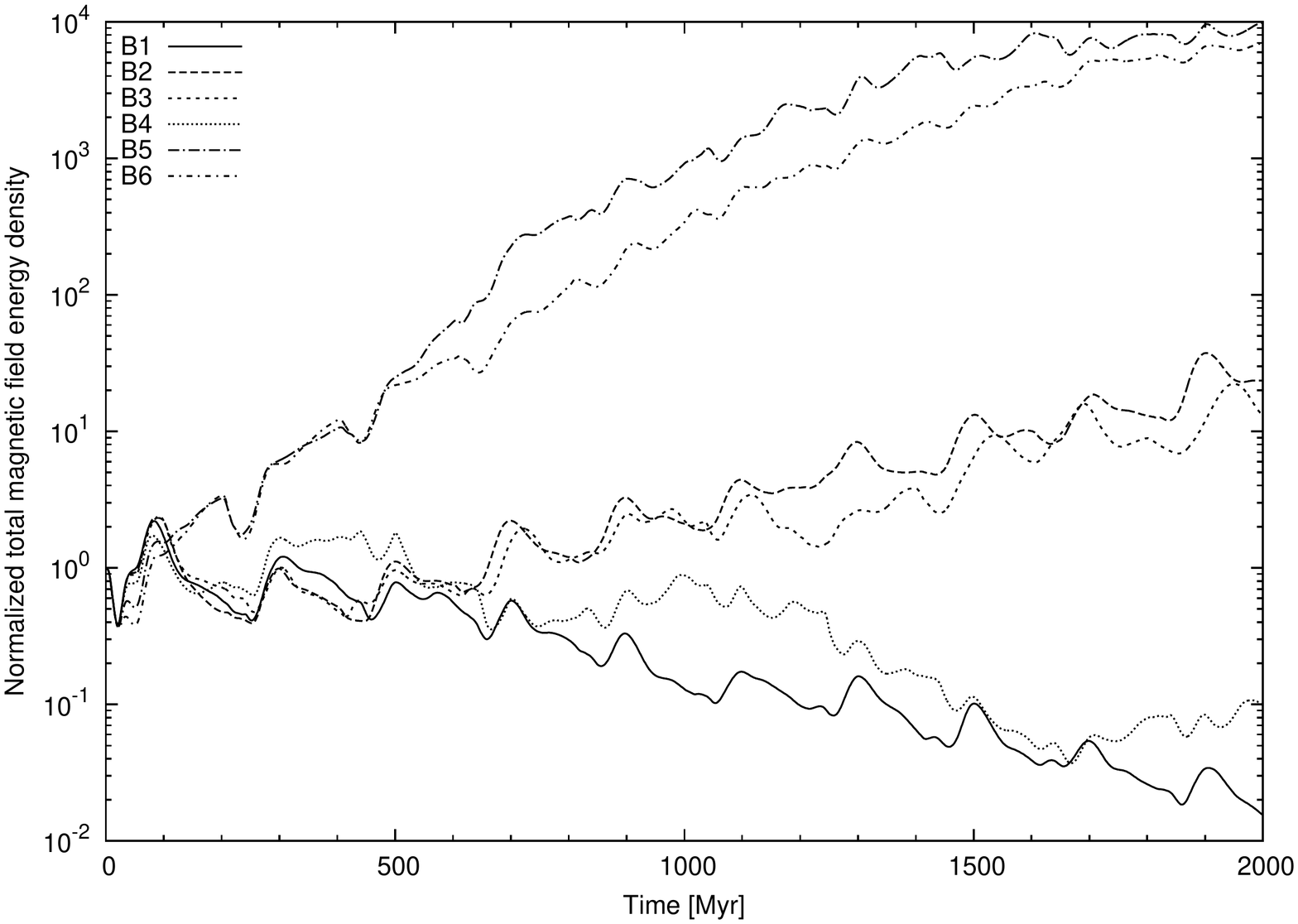}
\caption{Plots of total magnetic field energy evolution during 2~Myr. The left panel shows the evolution for
models (A1--A5) with different $\Omega$ (respectively 0.01--0.05~Myr$^{-1}$) and the right with different $q$:
B1--B3 for $q=0, 1, 1.5$ with $\Omega=0.01$~Myr$^{-1}$ and B4--B6 with $\Omega=0.05$~Myr$^{-1}$.\label{rys1}}
\end{figure}

We perform a two series of numerical experiments for different values of angular
velocity $\Omega$ and shearing parameter $q$ (see caption of Fig. \ref{rys1}).
In experiments A1--A5 we found that the efficiency of magnetic field
amplification depends strongly on the angular velocity (Fig. \ref{rys1} left),
but for $\Omega \geq 0.03$~Myr$^{-1}$ the amplification rate stabilizes and
becomes roughly independent on $\Omega$. In the second set of experiments we
change the shearing rate (Fig. \ref{rys1} right), for a constant angular
velocity. We find no amplification for experiments with $q=0$ (B1,B4). The other
simulations show that the dynamo process does not depend on shearing rate, as
long as $q$ is finite. We find a similar  evolution of the total magnetic field
energy for $q = 1$ and 1.5 (B2,B3 and B5,B6, respectively). 

To estimate the total production rate of magnetic field energy during simulation
time, we calculate the outflowing magnetic energy density through the $xy$ top
and bottom domain boundaries. Our results show that large amounts of magnetic
energy are lost through the open boundaries of the computational box, suggesting
that irregular galaxies due to their weaker gravitation can be efficient sources
of intergalactic magnetic fields.

\section{Conclusions}
In our work we found: a) the amplification of the total magnetic field energy in
irregular galaxies is possible even with slow rotation and a weak shear; b) for
dynamo action the shearing is needed, but the amplification rate on the shear is
weak; c) the larger angular velocity the higher efficiency of the dynamo
process; d) due to the weaker gravitation in irregular galaxies the outflow of
magnetic field off the simulation domain is large, suggesting that they may
effectively magnetize the intergalactic medium (Kronberg et al. 1999).

\begin{acknowledgments}
This work was supported by Polish Ministry of Science and Higher Education through grants:
92/N-ASTROSIM/2008/0, 2693/H03/2006/31 and 3033/B/H03/2008/35.
\end{acknowledgments}


\begin{thebibliography}{}


\bibitem[]{Chyzy00}
	{Chy\.{z}y, K.T., Beck, R., Kohle, S., Klein, U., Urbanik, M.} 2000,
	\textit{A\&A} 355, 128

\bibitem[]{Chyzy03}
	{Chy\.{z}y, K.T., Knapik, J., Bomans, D.J., Klein, U., Beck, R., Soida, M., Urbanik,~M.} 2003,
	\textit{A\&A} 405, 513

\bibitem[]{Ferriere98}
	{Ferri\`{e}re, K.} 1998,
	\textit{ApJ} 497, 759


\bibitem[]{Hanasz04}
	{Hanasz, M., Kowal, G., Otmianowska-Mazur, K., Lesch, H.} 2004,
	\textit{ApJ} (Letters) 605, L33

\bibitem[]{Hanasz06}
	{Hanasz, M., Kowal, G., Otmianowska-Mazur, K., Lesch, H.} 2006,
	\textit{AN} 327, 469


\bibitem[]{Hawley95}
	{Hawley, J.F., Gammie, C.F., Balbus, S.A.} 1995,
	\textit{ApJ} 440, 742

\bibitem[Kronberg et al. (1999)]{Kronberg99}
	{Kronberg, P. P., Lesch, H., Hopp, U.} 1999
	\textit{ApJ} 551, 56


\end{thebibliography}
\end{document}